\documentclass[a4paper, 12pt, oneside]{article}
\usepackage{jheppub}
\makeatletter
\def\@fpheader{\hfill\today}
\makeatother
\usepackage{amsmath,graphicx,color,epsfig,amssymb,float}
\usepackage[normalem]{ulem}
\usepackage{epstopdf}
\newcommand{\nc}{\newcommand}
\nc{\non}{\nonumber}
\nc{\hc}{\hbox {H.c.}}
\nc{\noi}{\noindent}
\nc{\barx}{\bar{x}}
\nc{\pbarn}{\;\hbox {pb}}
\nc{\fbarn}{\;\hbox {fb}}
\nc{\hsp}{\hspace{0.5cm}}
\nc{\lsp}{\hspace{1cm}}
\nc{\Lsp}{\hspace{2cm}}
\nc{\LLsp}{\hspace{3cm}}
\nc{\lra}{\longrightarrow}
\nc{\p}{\prime}
\nc{\sgn}{\text{sgn}}
\DeclareMathOperator{\sech}{sech}

\nc{\beq}{\begin{equation}}  \nc{\eeq}{\end{equation}}
\nc{\bea}{\begin{eqnarray}}  \nc{\eea}{\end{eqnarray}}
\nc{\baa}{\begin{array}}     \nc{\eaa}{\end{array}}
\nc{\bit}{\begin{itemize}}   \nc{\eit}{\end{itemize}}
\nc{\ben}{\begin{enumerate}} \nc{\een}{\end{enumerate}}
\nc{\bce}{\begin{center}}    \nc{\ece}{\end{center}}
\nc{\bpm}{\begin{pmatrix}}   \nc{\epm}{\end{pmatrix}}
\nc{\bvt}{\begin{verbatim}}  \nc{\evt}{\end{verbatim}}

\def\lsim{\mathrel{\raise.3ex\hbox{$<$\kern-.75em\lower1ex\hbox{$\sim$}}}}
\def\gsim{\mathrel{\raise.3ex\hbox{$>$\kern-.75em\lower1ex\hbox{$\sim$}}}}

\def\udots{\mathinner{\mkern1mu\raise1pt\vbox{\kern7pt\hbox{.}}\mkern2mu\raise4pt\hbox{.}\mkern2mu\raise7pt\hbox{.}\mkern1mu}}

\title{Thick-Brane Cosmology}
\author[a,b,\ast]{Aqeel Ahmed,\note[$\ast$]{On leave of absence from National Centre for Physics,
Quaid-i-Azam University Campus, 45320 Islamabad, Pakistan}}
\author[a]{Bohdan Grzadkowski}
\author[b]{and Jose Wudka}
\affiliation[a]{Faculty of Physics,
University of Warsaw,\\
Ho\.za 69, 00-681 Warsaw, Poland}
\affiliation[b]{Department of Physics and Astronomy, UC Riverside,\\
Riverside, CA 92521, USA}
\emailAdd{aqeel.ahmed@fuw.edu.pl}
\emailAdd{bohdan.grzadkowski@fuw.edu.pl}
\emailAdd{jose.wudka@ucr.edu}

\date{\today}
\abstract{
We search for time-dependent solutions for the 5-dimensional system of a scalar field canonically
coupled to gravity. Time-independent and time-dependent scalar field configurations with the most general
homogeneous and isotropic 4D metric are considered.
For the case of time-independent scalar field, the time evolution of the scale factor is obtained
for different values of the spatial curvature $k=0,\pm 1$.
In the case of time-dependent scalar field, two classes of solutions are discussed
and an extension of the superpotential formalism is proposed.
}
\arxivnumber{1312.3576}

\begin{document}
\maketitle
\flushbottom
\section{Introduction}
\label{Introduction}

The idea of extra space-time dimensions has received a lot of attention since the last decade or so as a
possible solution to the hierarchy problem \cite{ArkaniHamed:1998rs,Antoniadis:1998ig,Randall:1999ee}.
The  celebrated proposal by Randall and
Sundrum \cite{Randall:1999ee}, involves one extra-dimension with non-trivial warp factor
appearing due to the assumed anti-de Sitter (AdS) geometry along the fifth-dimension. This is
the Randall-Sundrum model with two D3-branes on the $S_1/Z_2$ orbifold along the extra-dimension,
which we refer to as RS1.
In this model the solution to the hierarchy problem is achieved by the virtue of the non-trivial
warped geometry
along the extra-dimension. It was also pointed out by Randall and Sundrum
that the extra-dimension
can be infinite and yet it can lead to nearly standard 4D gravity~\cite{Randall:1999vf};
we refer to this model as RS2, as is customary.
The main idea in RS2 is that a single D3-brane of positive tension
is embedded in a five-dimensional (5D) AdS geometry and the gravity is effectively four-dimensional (4D) at large
distances. The cosmological implications of the extra-dimensions, and in particular to that of RS models, have been
 studied in detail by many groups \cite{Binetruy:1999hy,Binetruy:1999ut,Cline:1999ts,Csaki:1999jh,Csaki:1999mp,
Flanagan:1999cu,Kanti:1999nz,
Kanti:1999sz,Bazeia:2007vx} that found that nearly 4D cosmological evolutions of our universe is recovered.

All the above-mentioned studies in brane cosmology were done in the presence of ``singular branes''.
The purpose of this paper is to study the cosmology of models where singular branes are replaced
by regularized counterparts generated by a 5D scalar field $\phi$ with
a non-trivial profile whose effects approach those of a singular brane
in an appropriate limit. There have been only few studies in this direction, where
the {\it smooth/thick brane} cosmological implications have been discussed \cite{Giovannini:2007xb,
George:2008vu,Kadosh:2012ru}. In this work we consider a scalar field with and without
time-dependence in the presence of 5D gravity. For the time-independent scalar field configurations,
the evolution of the scale factor are discussed and determined.
The time-dependent scalar field $\phi(t,y)$ case is discussed with certain assumptions on the
metric and an analogue of the traditional~\cite{DeWolfe:1999cp} superpotential method is developed.
It is found that for the case of time-dependent scalar field $\phi(t,y)$ there exists a class of solutions
that depend on time $t$ and the 5D coordinate $y$  only through the combination
$\eta=ct+dy$, where $c,d$ are constants.

The paper is organized as follows. Section~\ref{Introduction} presents an introduction and our
motivations. Our results are contained in Sec.~\ref{Thick brane cosmological solutions}, where
we separate the case of time independent (subsection \ref{Static thick brane solutions})
and time-dependent (subsection \ref{Time dependent thick brane solutions}) scalar field profile.
Conclusions are collected in sec.~\ref{conclusions} while the Appendix \ref{5D Dilaton-like solutions}
is dedicated to 5D dilaton-like solutions.

\section{Thick brane cosmological solutions}
\label{Thick brane cosmological solutions}
We will consider 5D space-times for which the metric takes
the following (4D conformal) form,
\begin{align}
ds^2&=a^2(t,y)g_{\mu\nu}dx^\mu dx^\nu+dy^2, \label{metric}
\end{align}
where $x^\mu$ are 4D coordinates while $g_{\mu\nu}$ is the 4D metric
that we take as the usual Robertson-Walker metric. The function $a(t,y)$ is a scale factor;
we will also refer to it as a warp factor because of its $y$.~\footnote{In our convention
capital Roman indices will refer to 5D objects,
i.e., $M,N,\cdots=0,1,2,3,5$ and the Greek indices label 4D objects, i.e., $\mu,\nu,\cdots=0,1,2,3$,
whereas, the lowercase Roman indices $i,j,\cdots=1,2,3$ represent the 3D spacial coordinates.}

The action for scalar field in the presence of 5D gravity is,
\begin{equation}
S=\int dx^5 \sqrt{-g}\left\{2M_\ast^{3}R-\frac{1}{2}g^{MN}\nabla_{M}\phi\nabla_{N}\phi-V(\phi)\right\},
\label{action}
\end{equation}
where $M_\ast$ is the Planck mass of the fundamental 5D theory and $R$ is the 5D Ricci scalar.
We assume that the scalar field $\phi$ depends exclusively on time and the extra coordinate $y$;
$V(\phi)$ is the potential for the scalar field.

The Einstein equation and the equation of motion for $\phi$ resulting from the above action \eqref{action} are
\begin{eqnarray}
R_{MN}-\frac{1}{2}g_{MN}R&=&\frac{1}{4M_\ast^{3}}T_{MN},\label{eineq1}\\
\nabla^{2}\phi-\frac{dV}{d\phi}&=&0, \label{eineq2}
\end{eqnarray}
where $\nabla^2$ is 5D covariant d'Alembertian operator and the energy-momentum tensor $T_{MN}$ for the
scalar field $\phi(t,y)$ is,
\begin{equation}
T_{MN}=\nabla_{M}\phi\nabla_{N}\phi-g_{MN}\left(\frac{1}{2}(\nabla\phi)^{2}+V(\phi)\right).
\label{emt}
\end{equation}
The explicit form of the components of the Einstein equation for the metric ansatz \eqref{metric}
can be written as~\footnote{For simplicity from here on we will consider the unit system such that $4M_\ast^{3}=1$.},
\begin{align}
00:&\lsp 3\left[\frac{1}{a^2}\frac{\dot a^2}{a^2}-\left(\frac{a^{\prime\prime}}{a} +\frac{a^{\p2}}{a^2}\right)+
\frac{k}{a^2}\right]= \frac{1}{2}\phi^{\prime2}+\frac{1}{2}\frac{1}{a^2}\dot\phi^2+V(\phi), \label{ein_00}\\
ij:&\lsp  \frac{1}{a^2}\left(2\frac{\ddot a}{a} -\frac{\dot a^2}{a^2}\right)-3\left(\frac{a^{\prime\prime}}{a}+
\frac{a^{\p2}}{a^2}\right)+\frac{k}{a^2}=
\frac{1}{2}\phi^{\prime2}-\frac{1}{2}\frac{1}{a^2}\dot\phi^2+V(\phi), \label{ein_ij}\\
05:&\lsp \frac{a^\prime}{a}\frac{\dot a}{a}-\frac{\dot a^\prime}{a}= \frac{1}{3}\phi^{\prime}\dot\phi,
\label{ein_05}\\
55:&\lsp 3\left[2\frac{ a^{\p2}}{a^2}-\frac{1}{a^2}\frac{\ddot a}{a} -\frac{k}{a^2}\right] =\frac{1}{2}\phi^{\prime2}+\frac{1}{2}\frac{1}{a^2}\dot\phi^2-V(\phi).
\label{ein_55}
\end{align}
where $k = 0,~\pm1$
denotes the spatial curvature of the 4D homogeneous and isotropic
space-time for Minkowski, de Sitter and anti-de Sitter space, respectively.

The scalar field equation of motion can be written as,
\beq
\phi^{\p\p}-\frac{1}{a^2}\ddot\phi+4\frac{a^\p}{a}\phi^\p -\frac{2}{a^2}\frac{\dot a}{a}\dot\phi-\frac{dV}{d\phi}=0. \label{phi_eom}
\eeq
In our notation a prime (dot) denotes a $y$ ($t$) derivative.

In the following two subsections we will consider two cases, one with time-independent (static) scalar
field and the other with time-dependent scalar field.

\subsection{Static thick brane solutions}
\label{Static thick brane solutions}
In this subsection we will consider a static scalar field scenario, in other words we assume  $\phi(t,y)=\phi(y)$,
 but still allow $a$ to be time-dependent.
In this case the Einstein equations \eqref{ein_00}-\eqref{ein_55}
simplify as follows,
\begin{align}
00:&\lsp  3\left[\frac{1}{a^2}\frac{\dot a^2}{a^2}-\left(\frac{a^{\prime\prime}}{a} +\frac{a^{\p2}}{a^2}\right)+\frac{k}{a^2}\right]=\frac{1}{2}\phi^{\prime2}+V(\phi), \label{ein_00_s}\\
ij:&\lsp  \frac{1}{a^2}\left(2\frac{\ddot a}{a} -\frac{\dot a^2}{a^2}\right)-3\left(\frac{a^{\prime\prime}}{a}+\frac{a^{\p2}}{a^2}\right)+\frac{k}{a^2}=
\frac{1}{2}\phi^{\prime2}+V(\phi), \label{ein_ij_s}\\
05:&\lsp  \frac{a^\prime}{a}\frac{\dot a}{a}-\frac{\dot a^\prime}{a}=0, \label{ein_05_s}\\
55:&\lsp  3\left[2\frac{ a^{\p2}}{a^2}-\frac{1}{a^2}\frac{\ddot a}{a} -\frac{k}{a^2}\right]  =
\frac{1}{2}\phi^{\prime2}-V(\phi). \label{ein_55_s}
\end{align}
The equation of motion of the scalar field \eqref{eineq2} has the following form,
\beq
\phi^{\p\p}+4\frac{a^\p}{a}\phi^\p -\frac{dV}{d\phi}=0.
\label{phi_eom_s}
\eeq

\subsubsection{Evolution of the scale factor}

The assumption that scalar field $\phi$ is time independent implies that $T_{05}=0$,
consequentially $G_{05}=0$. This requires $ \partial_t \partial_y \ln a =0 $,
which implies that $a$ is separable:  $a(t,y)={\hat a}(t)\bar a(y)$.
Using this the remaining Einstein equations become
\begin{align}
00:&\lsp  \frac{1}{{\hat a}^2}\frac{\dot {{\hat a}}^2}{{\hat a}^2}+\frac{k}{{\hat a}^2}=
\frac{\bar a^2}{3}\left[3\left(\frac{\bar a^{\prime\prime}}{\bar a} +\frac{\bar a^{\p2}}{\bar a^2}\right)+
\frac{1}{2}\phi^{\prime2}+V(\phi)\right], \label{ein_00_a}\\
ij:&\lsp  \frac{1}{{\hat a}^2}\left(2\frac{\ddot {\hat a}}{{\hat a}} -
\frac{\dot {\hat a}^2}{{\hat a}^2}\right)+\frac{k}{{\hat a}^2}=\bar a^2\left[3\left(\frac{\bar a^{\prime\prime}}{\bar a}+
\frac{\bar a^{\p2}}{\bar a^2}\right)+\frac{1}{2}\phi^{\prime2}+V(\phi)\right], \label{ein_ij_a}\\
55:&\lsp  \frac{1}{{\hat a}^2}\frac{\ddot {\hat a}}{{\hat a}} +\frac{k}{{\hat a}^2} =
\frac{\bar a^2}{3}\left[6\frac{ \bar a^{\p2}}{\bar a^2}-\frac{1}{2}\phi^{\prime2}+V(\phi)\right]. \label{ein_55_a}
\end{align}
where the left(right) hand sides depend only on $t$ ($y$).
We then obtain the following set of equations for $\hat a(t)$:
\begin{align}
00:&\lsp  \frac{1}{{\hat a}^2}\frac{\dot {{\hat a}}^2}{{\hat a}^2}+\frac{k}{{\hat a}^2}=C_{t},
\label{ein_00_a_t}\\
ij:&\lsp  \frac{1}{{\hat a}^2}\left(2\frac{\ddot {\hat a}}{{\hat a}} -
\frac{\dot {\hat a}^2}{{\hat a}^2}\right)+\frac{k}{{\hat a}^2} = C_{x},
\label{ein_ij_a_t}\\
55:&\lsp  \frac{1}{{\hat a}^2}\frac{\ddot {\hat a}}{{\hat a}} +\frac{k}{{\hat a}^2} = C_{y},
\label{ein_55_a_t}
\end{align}
where $C_{t,x,y}$ are constants. It is easy to see that in order for the first two equations to be
consistent with the third one it is necessary that
\beq
C_{y}=\frac{C_{t}+C_{x}}{2}.
\label{Lam_55}
\eeq
On the other hand, form the right hand sides of \eqref{ein_00_a}-\eqref{ein_55_a} one obtains
for the $y$-dependent functions the following equations
\begin{align}
00:&\lsp C_{t}=\frac{\bar a^2}{3}\left[3\left(\frac{\bar a^{\prime\prime}}{\bar a} +\frac{\bar a^{\p2}}{\bar a^2}\right)+\frac{1}{2}\phi^{\prime2}+V(\phi)\right], \label{ein_00_a_y}\\
ij:&\lsp C_{x}=\bar a^2\left[3\left(\frac{\bar a^{\prime\prime}}{\bar a}+\frac{\bar a^{\p2}}{\bar a^2}\right)+\frac{1}{2}\phi^{\prime2}+V(\phi)\right], \label{ein_ij_a_y}\\
55:&\lsp C_{y}=\frac{\bar a^2}{3}\left[6\frac{ \bar a^{\p2}}{\bar a^2}-\frac{1}{2}\phi^{\prime2}+V(\phi)\right].
\label{ein_55_a_y}
\end{align}
Then \eqref{ein_00_a_y}-\eqref{ein_ij_a_y} immediately imply that
\beq
C_{x}= 3 C_{t},
\label{Lam_ij}
\eeq
so that all the constants can be expressed in terms $C_{y}$ that will be denoted by $\bar\Lambda$:
\beq
C_y = \bar\Lambda, \lsp C_{t}=\frac12 \bar\Lambda, \lsp C_{x}= \frac32 \bar\Lambda.
\label{Lam_sol}
\eeq
Then one finds the following equations that must be satisfied by $\hat a(t)$:
\bea
\frac{\ddot {\hat a}}{{\hat a}} -2\frac{\dot {\hat a}^2}{{\hat a}^2}-k&=&0,   \label{alpha_1}\\
\frac{\dot {{\hat a}}^2}{{\hat a}^2}-\frac{\bar\Lambda}{2}{\hat a}^2+k&=&0. \label{alpha_2}
\eea
As it will be shown below, even though the above equations look independent, solutions that satisfy both
of them do exist for each possible $k=0,\pm 1$.
Before we proceed, it is worth discussing the relations between \eqref{alpha_1} and \eqref{alpha_2}.
First note that Eq.~\eqref{alpha_2} can be rewritten as
\beq
\frac{\dot {{\hat a}}^2}{{\hat a}^4}-\frac{\bar\Lambda}{2}+\frac{k}{{\hat a}^2}=0, \label{alpha_3}
\eeq
which is a first integral of Eq.~\eqref{alpha_1}:
a derivative of \eqref{alpha_3} reproduces Eq.~\eqref{alpha_1}.
So, one can  recognize in \eqref{alpha_1} and \eqref{alpha_2} the analogs of classical
equation of motion and energy conservation, respectively. Therefore the role of Eq.~\eqref{alpha_2}
is just to adjust ``velocity'' at the initial moment such that
``energy'' is properly matched.

Note also that \eqref{alpha_2} is identical with the standard Friedman equation
written in terms of the conformal time, thus one finds that $\bar\Lambda$ has the interpretation
of 4D cosmological constant.

\begin{figure}[!hbt]
\centering
\begin{tabular}{cc}
\includegraphics[scale=0.62]{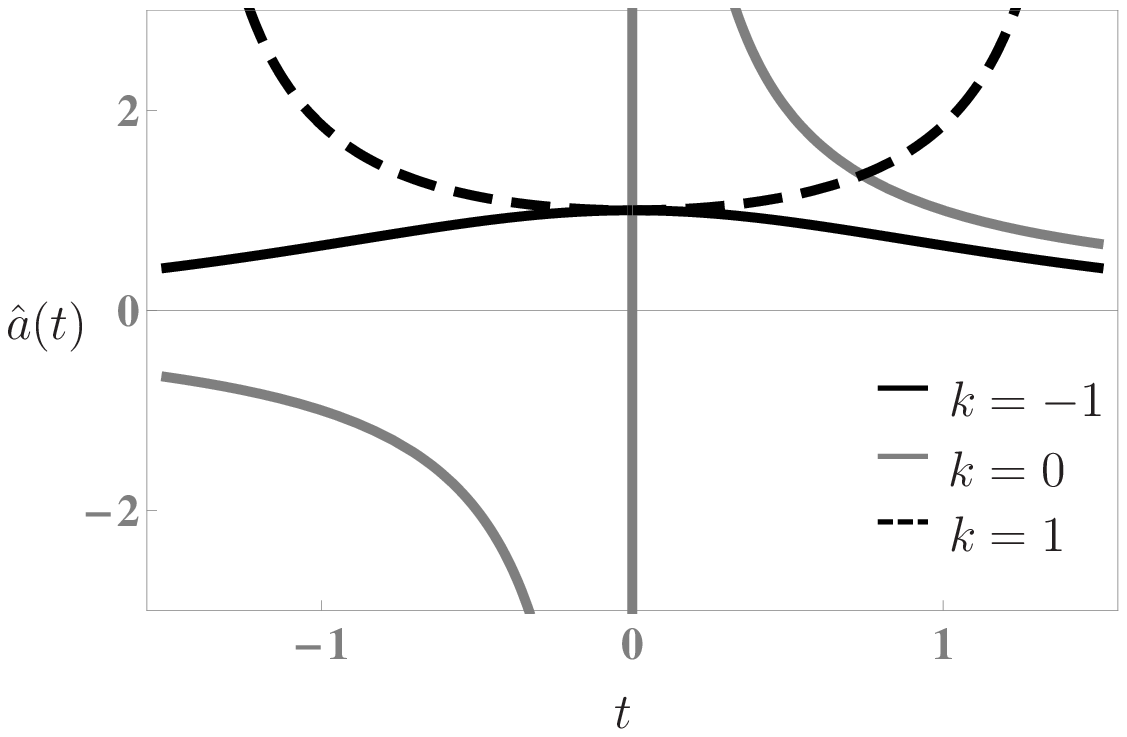} & \includegraphics[scale=0.62]{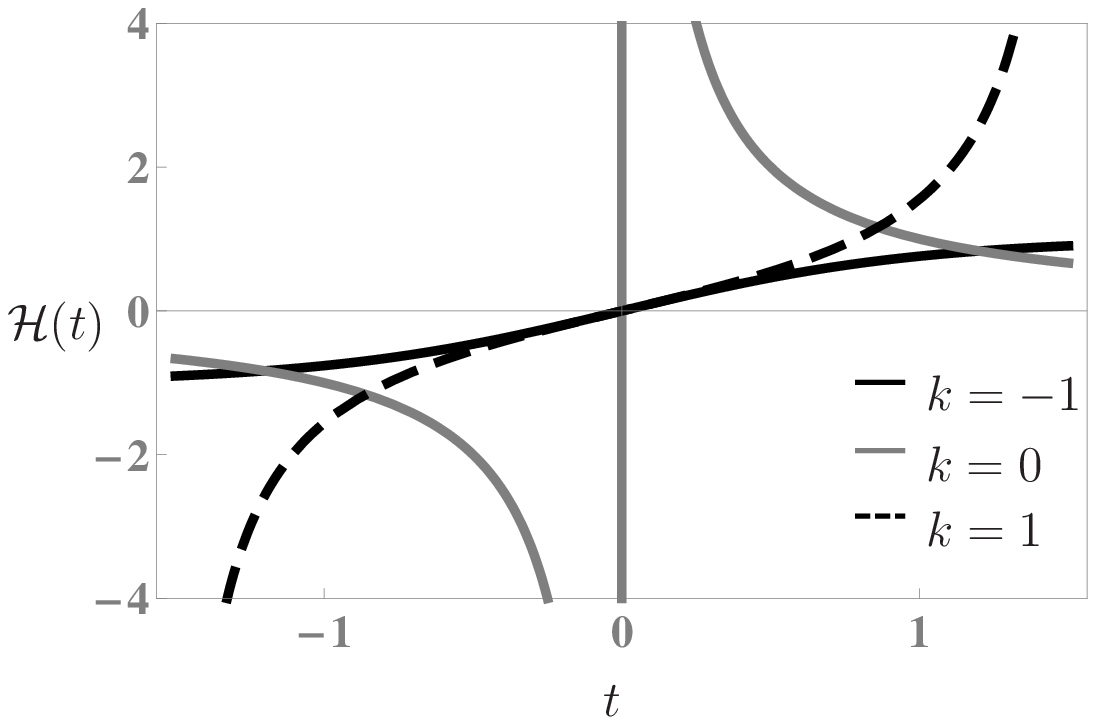}
\end{tabular}
\caption{The left graph shows the behavior of the 4D conformal scale factor ${\hat a}(t)$ as a
function of $t$, whereas, the right graph presents the Hubble parameter ${\cal H}(t)$ as a
function of $t$ for different values of spacial curvature $k$. We choose the value of
constants $\alpha_0=0$ and $\alpha_1=\sqrt2$.}
\label{fig_alpha}
\end{figure}

The solutions of Eq. \eqref{alpha_1} are
\beq
{\hat a}(t) =|2/\bar\Lambda|^{1/2}\left\{\baa{ccc}
\sech(t+\alpha_0) &\lsp  k=-1 & \quad (\bar\Lambda<0)\\
1/(t+\alpha_0) &\lsp  k=0 & \quad (\bar\Lambda>0)\\
\sec(t+\alpha_0)&\lsp  k=+1 & \quad (\bar\Lambda>0)
\eaa
\right.,
\label{alpha_k0}
\eeq
where $\alpha_0$ is an integration constant.
Note that for $k=0,\,1$ the metric is singular at a finite time $t_{\rm sing}=-\alpha_0 + (n+1/2)\pi k$,
where $n$ is an integer. Using (\ref{alpha_k0}) it is easy to determine the evolution of the 4D conformal Hubble parameter
${\cal H}\equiv\dot {\hat a}(t)/{\hat a}(t)$
\beq
{\cal H}(t)=\left\{\baa{ccc}
\tanh(t+\alpha_0) &\lsp  k=-1 & \quad (\bar\Lambda<0)\\
1/(t+\alpha_0) &\lsp  k= 0 & \quad (\bar\Lambda>0)\\
\tan(t+\alpha_0)&\lsp  k=+1 & \quad (\bar\Lambda>0)
\eaa
\right..
\label{Hubble_k0}
\eeq
Note that when $k=-1$, $\bar\Lambda$ must be negative,
which corresponds to  anti-de Sitter geometry;
for $k=0, +1$, $\bar\Lambda$ must be positive, thus representing  de Sitter space-time.
In Fig. \ref{fig_alpha} we have plotted the scale factor $\hat a(t)$ and the Hubble parameter
${\cal H}(t)$ for $k=0,~\pm 1$.

When $ \bar\Lambda =0 $ there are no (real) solutions when $ k = + 1 $.
When $k =-1$, known as the Milne universe in the conventional cosmology,
Eqs.~\eqref{ein_ij_a_y}-\eqref{ein_55_a_y}
reduce to the standard static equations considered e.g. in \cite{DeWolfe:1999cp,Ahmed:2012nh}.
In this case
time-dependent part of the
scale factor is  determined by Eqs.~\eqref{alpha_1}-\eqref{alpha_2}, whose general solution is
a linear combination of the following functions
\beq
\hat a_\pm(t) = \alpha_\pm e^{\pm \sqrt{-k} t}, \lsp  k=-1 ,\, 0.
\label{a_sol}
\eeq
Note  that the static solution requires $k=0$ and  $\bar\Lambda=0$.

In the following section we focus on the $y$-dependent solutions of Eqs.~\eqref{ein_ij_a_y} and \eqref{ein_55_a_y}.

\subsubsection{Extra-dimensional profiles}

In this section we will determine $y$-dependent part of solutions that are governed by
Eqs.~\eqref{ein_ij_a_y}-\eqref{ein_55_a_y}. For this purpose
it is useful to define $\bar a(y)\equiv e^{A(y)}$, such that our $y$-dependent Einstein equations
Eqs.~\eqref{ein_ij_a_y}-\eqref{ein_55_a_y} and the scalar field equation \eqref{phi_eom_s} can be written as,
\begin{align}
3A^{\p\p}+\frac32\bar\Lambda e^{-2A}&=-\phi^{\p2}, \label{ein_A1}\\
6A^{\p2}-3\bar\Lambda e^{-2A} &=\frac{1}{2}\phi^{\prime2}-V(\phi),   \label{ein_A2}\\
\phi^{\p\p}+4A^{\p}\phi^\p -\frac{dV}{d\phi}&=0. \label{phi_eom_A}
\end{align}
The procedure we follows begins by assuming
$A(y)$ is a known function, so the above conditions are to be considered
as equations to determine $\phi(y)$ and $V(\phi)$.~\footnote{In the literature there are few known analytic de Sitter
and anti-de Sitter solutions of the system \eqref{ein_A1}-\eqref{ein_A2}, see for example
\cite{Gremm:2000dj,Kobayashi:2001jd,Wang:2002pka,Sasakura:2002tq,Afonso:2006gi}.}.

\begin{figure}[!hbt]
\centering
\includegraphics[scale=0.65]{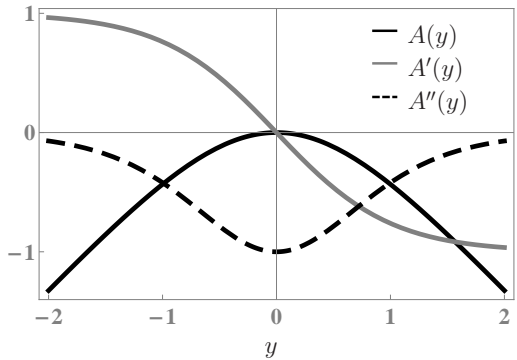} 
\caption{The warp function $A(y)$ and its derivatives $A^\p(y)$ and $A^{\p\p}(y)$ as a function of $y$ for $\beta=1$.}
\label{fig_A}
\end{figure}

Specifically, we will consider the following form of the warp function $A(y)$,
\beq
A(y)=-\ln\cosh(\beta y),    \label{A_y}
\eeq
where $\beta$ is a parameter. The above choice is dictated by simplicity and by a desire
to have a warp factor which behaves as $\sim  \exp{(-|y|)}$ at large $y$, so that
it mimics RS solutions and the hierarchy problem can be in principle approached.
This choice of $A(y)$ approximates well the static solution obtained e.g. in \cite{DeWolfe:1999cp,Ahmed:2012nh} for
a kink profile of the scalar field. Figure~\ref{fig_A} shows the warp function and its derivatives.

\begin{figure}[!hbt]
\centering
\begin{tabular}{cc}
\includegraphics[scale=0.62]{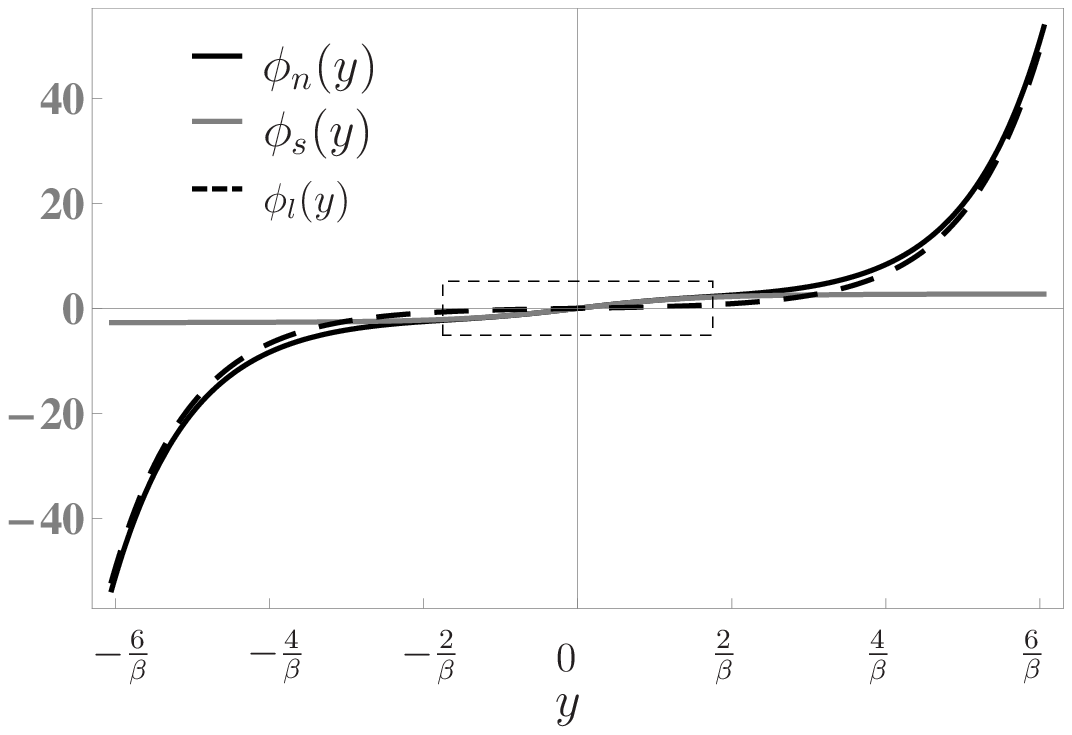} & \includegraphics[scale=0.62]{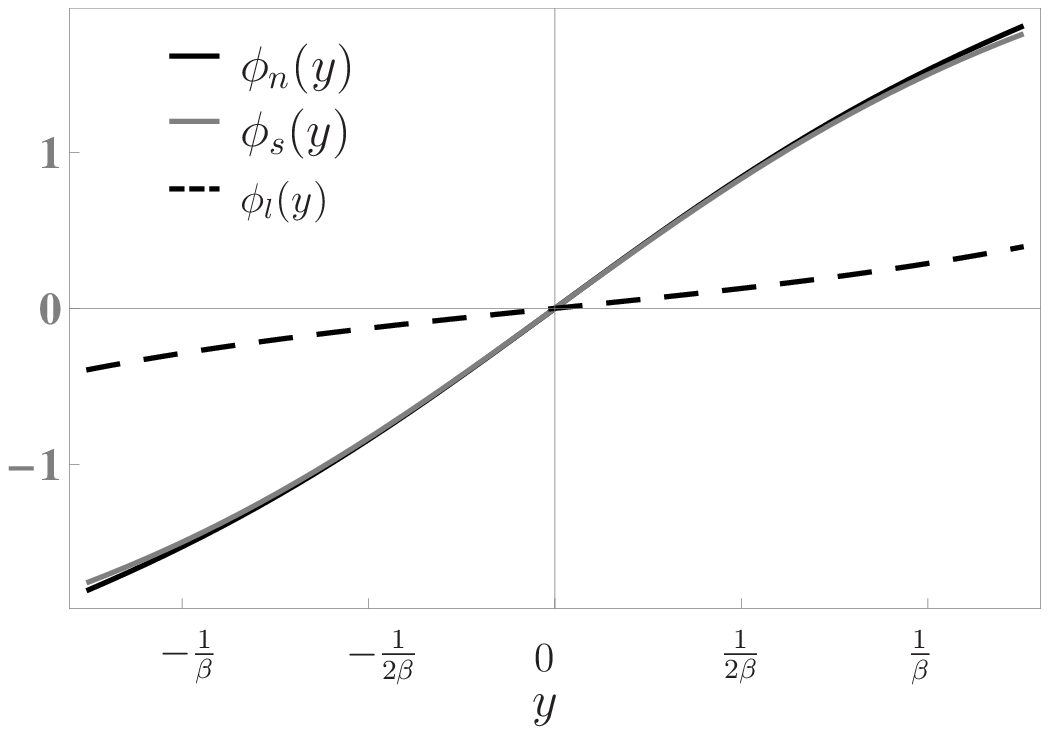}
\end{tabular}
\caption{These graphs show the exact numerical solution for the scalar field $\phi_n(y)$,
the approximate analytic solution for small ($\phi_s(y)$) and large ($\phi_l(y)$) values of $y$
as a function of $y$ in units of $\beta^{-1}$ for $\beta=5$ and $\bar\Lambda=-1$.
The right graph shows the zoomed central region of the left graph.}
\label{fig_phi}
\end{figure}

It is difficult to find an exact analytical solution for $\phi(y)$ from Eq. \eqref{ein_A1}, however
if one considers separately regions of small ($|y|\lsim \beta^{-1}$) and large ($|y|\gsim \beta^{-1}$)
$y$ then approximate analytical solutions are easy to obtain.
When $\beta$ is large (which is the case of our interest) then for small values of $y$, one can ignore~\footnote{
When $y\to 0$ then $A(y)\to 0$ and $A^{\p\p}\to -\beta^2$, therefore in the vicinity of $y=0$
the Eq.~\eqref{ein_A1} behaves as $-3\beta^2+\frac32\bar\Lambda=-\phi^{\p2}$, implying $\phi_s(y)$ linear
in $y$. For values of $\beta$ adopted here the $\bar\Lambda$ term is negligible.}
the exponential term in Eq.~\eqref{ein_A1}, i.e.
\beq
3A^{\p\p}=-\phi^{\p2}, \quad (y \to 0 ) \label{ein_A1_a}
\eeq
with the following solution
\beq
\phi_s(y)=2\sqrt3\arctan[\tanh(\beta y/2)], \quad (y \to 0 )
\label{phi_s}
\eeq
where $\phi_s(y)$ denotes the solution for small $y$. It is important to note that dropping the exponential term in
Eq. \eqref{ein_A1} is a reasonable assumption in the vicinity of $y=0$ if $\beta$ is larger than $\bar\Lambda$, as
illustrated in Fig.~\ref{fig_phi} for $\beta=5$ and $\bar\Lambda=-1$.
On the other hand, for large values of $y$, the exponential term dominates in Eq. \eqref{ein_A1}
and we can ignore $A^{\p\p}(y)$, i.e.,
\beq
\frac32\bar\Lambda e^{-2A}=-\phi^{\p2},   \label{ein_A1_b}
\eeq
with $\bar\Lambda < 0$. The solution of above equation reads
\beq
\phi_l(y)=\sqrt{-\frac{3\bar\Lambda}2}\frac{1}{\beta}\sinh(\beta y),
\quad (|y| \to \infty) \label{phi_l}
\eeq
where $\phi_l(y)$ denotes the solution valid for large values of $y$.

In Fig. \ref{fig_phi} we have plotted the approximate analytic solutions $\phi_{s,l}(y)$
and the exact numerical one $\phi_n(y)$. For large $y$ the
quality of the approximation can be easily estimated
from the figure; one finds that for $|y| \gsim 5\beta^{-1}$, $\phi_n\simeq \phi_l$.
For small $y$ the right panel of the figure
shows that for $|y| \lsim \beta^{-1}$, $\phi_n\simeq \phi_s$. In the intermediate region
$\beta^{-1}\lsim |y| \lsim 5 \beta^{-1}$ the approximations
$\phi_{n,s}$ are less accurate.
It is also worth to mention that as $\beta$  grows the region of applicability
of $\phi_s$ shrinks, and $\phi_l$ converges to the
exact numerical solution $\phi_n$.

There is a comment here in order. If, instead of \eqref{A_y}, we had used the solution
obtained in the static case of \cite{Ahmed:2012nh}, then for $\bar\Lambda=0$
we would reproduce exactly the kink profile for the scalar field and the corresponding potential
as in \cite{Ahmed:2012nh}. In that case (with $k=-1$), the time evolution of the scale factor
would be governed by \eqref{a_sol} while the scalar profile would preserve its shape. In this special case
the time evolution in 4D and source (the scalar field profile) along the extra dimension fully decouple,
so that the scalar profile is retained while non-trivial time evolution of the scale factor
has purely 4D nature.

\begin{figure}[!hbt]
\centering
\includegraphics[scale=0.75]{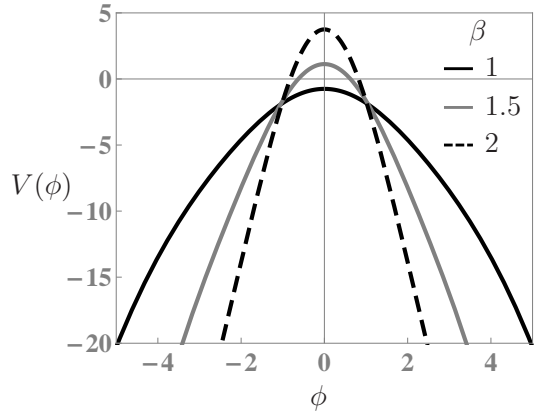} 
\caption{The scalar potential $V(\phi)$ as a function of $\phi$ for different values of $\beta$ and $\bar\Lambda=-1$.}
\label{fig_Vphi}
\end{figure}

We then substitute the solutions we obtained for $A(y)$ and $\phi(y)$ in \eqref{ein_A2} to obtain the scalar
potential $V(\phi)$, which we plot as a function of $\phi$ in Fig. \ref{fig_Vphi}.
To get approximate analytic results for the scalar potential $V(\phi)$ corresponding to
small and large values of $\phi(y)$, we use the Einstein equation \eqref{ein_A2} along with
the analytic solutions of scalar field $\phi_s(y)$ and $\phi_l(y)$.
For small values of scalar field $\phi_s(y)$ the scalar potential is~\footnote{For $\bar\Lambda=0$
one would reproduce
the standard bottom of a wine bottle potential as in \cite{Ahmed:2012nh}.},
\beq
V_s(\phi)
=\left(\frac32\beta^2+\frac94\bar\Lambda\right)+\left(-\frac52\beta^2+\frac34\bar\Lambda\right)\phi^2+
{\cal{O}}\left(\phi^4\right),  \quad\quad (\phi \to 0 ). \label{V_s}
\eeq

For large values of scalar field $\phi_l(y)$ we can write the potential as,
\beq
V_l(\phi)= -\frac32\beta^2\phi^2+\left(-6\beta^2+\frac94 \bar\Lambda\right) +{\cal{O}}\left(\phi^{-2}\right),
\quad\quad (|\phi| \to \infty ). \label{V_l}
\eeq

Note that the potential is unbounded from below. In fact this is a generic consequence of
the requirement that the warp function $A(y)$ has a linear
dependence on $y$ as $|y| \to \infty $.
Indeed, as seen from \eqref{ein_A1}-\eqref{ein_A2}, for large $y$,
$V(y) \sim \frac94\bar\Lambda e^{-2A} \sim \frac94\bar\Lambda e^{2\beta|y|}$.
Since at large $|y|$, $\phi \propto e^{\beta |y|}$  we find
$V(\phi) \propto -\phi^2$, so the potential is always unbounded from below.

An alternative approach to solve Eqs. \eqref{ein_A1}-\eqref{ein_A2} is to reduce these equations into
the following first order equations by the use of an extension of the superpotential method \cite{DeWolfe:1999cp},
\begin{align}
A^\p&=-\frac{1}{3}{\cal W}\gamma(y),    \label{sp_A}\\
\phi^\p&=\frac{1}{\gamma(y)}\frac{\partial {\cal W}}{\partial \phi},   \label{sp_phi}\\
V&=\frac{1}{\gamma(y)^2}\left(\frac{\partial {\cal W}}{\partial \phi}\right)^2-\frac{2}{3}{\cal W}^2,
\label{sp_V}
\end{align}
where $\gamma(y)$ is defined by,
\beq
\gamma(y)=\left(1+\frac92\frac{\bar\Lambda}{{\cal W}^2}e^{-2A(y)}\right)^{1/2}. \label{gamma}
\eeq
The superpotential method  reduces the second order non-linear
differential equations \eqref{ein_A1}-\eqref{phi_eom_A} to the system of first order nonlinear
differential equations \eqref{sp_A}-\eqref{sp_V}, though  obtaining the solution are less straightforward
than when $\bar \Lambda=0$ (for $\bar\Lambda=0$ see, for example,
\cite{DeWolfe:1999cp,Ahmed:2012nh}).
Unfortunately, for the present case, one can not start with a desired shape for the scalar profile, our strategy is instead
to solve a system of first order nonlinear equations by first
choosing the warp function $A(y)$, next solve Eq. \eqref{sp_A} algebraically for ${\cal W}(y)$,
and then solve the following equation for $\phi(y)$
\beq
\phi^\p(y)=\sqrt{\frac{{\cal W}^\p(y)}{\gamma(y)}}.    \label{phi_p}
\eeq
Then Eq. \eqref{sp_V} gives the potential $V(y)$ which can be written as $V(\phi)$ after inverting $\phi(y)$ to $y(\phi)$.
We will not further investigate these solutions for $\phi(y)$,
we only note that choosing $A(y)$ as in \eqref{A_y} one would reproduce the result obtained earlier in this subsection.

\subsection{Time-dependent thick brane solutions}
\label{Time dependent thick brane solutions}

In this subsection we will look for solutions of the Einstein equations \eqref{ein_00}-\eqref{ein_55} allowing for
time-dependence of the scalar field.

\subsubsection{Boosted solutions}
\label{Boosted solutions}

In this subsection we show how the a static solution for the warp factor, $a(y)$ and scalar field $\phi(y)$
can be promoted to a time-dependent solution through a boost along the extra dimension: $y\to y'=\gamma(vt+y)$,
where $\gamma=1/\sqrt{1-v^2}$ and $v$ is a relative velocity.
It proves to be more convenient first to redefine the fifth coordinate $y$ so that the length element \eqref{metric}
is written as
\beq
ds^2=a^2(z)\left(\eta_{\mu\nu}dx^\mu dx^\nu+dz^2\right),
\label{con_metric}
\eeq
Let us consider a Lorentz transformation $t'=\gamma(t+v z)$ and $z'=\gamma (v t + z)$. It is easy to check that
$a'(t',z')=a(t,z)$, since $\phi$ is a scalar field $\phi'(t',z')=\phi(t,z)$. By general covariance  $a'(t',z')$ and $\phi'(t',z')$
are also solutions on the Einstein equations. Therefore we conclude that for any given stationary solution
$a(y)$ and $\phi(y)$, the functions $a[\gamma (-v t+z(y))]$ and $\phi[\gamma (-v t+z(y))]$
also satisfy the Einstein equations.
This strategy could be applied to any stationary solution, e.g. to the kink solution discussed in \cite{Ahmed:2012nh}.

\subsubsection{Twisted solutions}
\label{Twisted solutions}

In this subsection we return to the fifth dimensional coordinate $y$.
We will to show that one can obtain a class of interesting solutions assuming that $a$ and
$ \phi$ depend on $y$ and $t$ only through the combination
$\eta\equiv c t+ d y$~\footnote{Note that this is not a boost of a stationary solution.}
where $c$ and $d$ are non-zero constants. In the next subsection we will show that if the superpotential method
is used, the $05$ component of the Einstein equations
in fact implies such a dependence on $\eta$ for $\phi$.
With this assumption the Einstein equations \eqref{ein_00}-\eqref{ein_55} become,
\begin{align}
00:&\hsp 3\frac{c^2}{a^2}\frac{a^{\p2}}{a^2}-3d^2\left(\frac{a^{\prime\prime}}{a} +\frac{a^{\p2}}{a^2}\right)+\frac{3 k}{a^2}= \frac{d^2}{2}\phi^{\prime2}+\frac{1}{2}\frac{c^2}{a^2}\phi^{\p2}+V(\phi), \label{ein_00_B}\\
ij:&\hsp \frac{c^2}{a^2}\left(2\frac{a^{\p\p}}{a} -\frac{a^{\p2}}{a^2}\right)-3d^2\left(\frac{a^{\prime\prime}}{a}+
\frac{a^{\p2}}{a^2}\right)+\frac{k}{a^2}=\frac{d^2}{2}\phi^{\prime2}-\frac{1}{2}\frac{c^2 }{a^2}\phi^{\p2}+V(\phi), \label{ein_ij_B}\\
05:&\hsp \frac{a^{\p\p}}{a}-\frac{a^{\p2}}{a^2}= -\frac{1}{3}\phi^{\prime2},
\label{ein_05_B}\\
55:&\hsp 6d^2\frac{ a^{\p2}}{a^2}-3\frac{c^2}{a^2}\frac{a^{\p\p}}{a}-\frac{3 k}{a^2} =\frac{d^2}{2}\phi^{\prime2}+\frac{1}{2}\frac{c^2}{a^2}\phi^{\p2}-V(\phi).
\label{ein_55_B}
\end{align}
Where now a {\it prime} denotes an $\eta$ derivative.
If now we add Eqs.~\eqref{ein_ij_B}-\eqref{ein_55_B},
and then use of Eq. \eqref{ein_05_B} we obtain
\beq
\frac{a^{\p\p}}{a} +\frac{a^{\p2}}{a^2}+\frac{2k}{c^2}=0.
\label{s1}
\eeq
On the other hand, subtracting Eqs.~\eqref{ein_00_B}-\eqref{ein_ij_B} and using \eqref{s1} gives
\beq
\frac{a^{\p2}}{a^2}+\frac{k}{c^2}=\frac16\phi^{\p2}.
\label{s2}
\eeq
Using the above two relations in Eq. \eqref{ein_55_B}, we obtain the following form for the scalar potential $V(\phi)$
\beq
V(\phi)=d^2\left(-\frac12 \phi^{\p2}+\frac{6k}{c^2}\right). \label{potential_phiz}
\eeq
At this point the strategy is clear, one first solves \eqref{s1} for $a(\eta)$, then $\phi$ is easily
determined from  \eqref{s2}. If $\phi(\eta)$ is an invertible function of $\eta$
then  $V(\phi)$ can be found from \eqref{potential_phiz}. In the following we will find such
solutions for each possible value of $k$.

\paragraph{\textbf{$k=0:$}} In this case the warp factor $a(\eta)$
is obtained by integrating Eq. \eqref{s1}:
\beq
a(\eta)=a_0\left(1+2b_0\eta\right)^{1/2},     \label{az_sol}
\eeq
where $a_0$ and $b_0$ are integration constants. It is important to note that the above solution
is only valid in the region of the space-time where $\eta>-1/2b_0$, and
we will see below that the scalar field
is singular at $\eta\to-1/2b_0$. With this explicit expression for $a(\eta)$, we  use Eq.~\eqref{s2}
to find the scalar field $\phi(\eta)$:
\beq
\phi(\eta)=\pm\sqrt\frac32 \ln\left(1+2b_0\eta\right)+\phi_0,   \label{phiz_sol}
\eeq
where $\phi_0$ is an integration constant. Then the scalar potential
takes the form
\beq
V(\phi)=-3b_0^2e^{-\sqrt\frac83(\phi-\phi_0)}.
\label{potential_phi0}
\eeq
It is noteworthy that the above potential is similar to the dilaton potential
$V_{\rm dilaton}=-|\Lambda|e^{\sqrt{4/3} \phi}$
studied in a 5D context e.g. in \cite{Antoniadis:2011qw}; in our case,
however, the argument of the exponent is $\sqrt{8/3}\,\phi$, while in \cite{Antoniadis:2011qw}
it is $\sqrt{(4/3)}\phi$. In the Appendix~\ref{5D Dilaton-like solutions} we discuss this issue in details.

\begin{figure}[!hbt]
\centering
\includegraphics[scale=0.62]{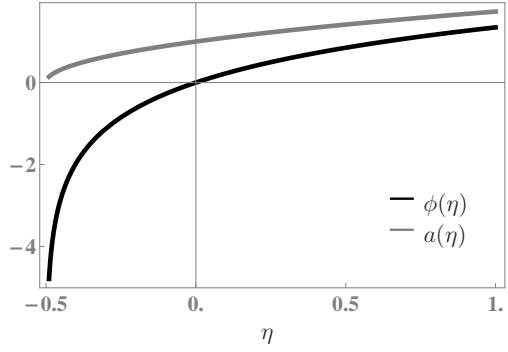}
\caption{This plot shows the behavior of $\phi(\eta)$ \eqref{phiz_sol} and $a(\eta)$ \eqref{az_sol}
as a function of $\eta$ for $k=0$, $a_0=b_0=1$ and $\phi_0=0$.}
\label{fig_K0}
\end{figure}

Figure \ref{fig_K0} shows the behavior of the scalar field $\phi(\eta)$ and the warp factor $a(\eta)$.

\begin{figure}[!hbt]
\centering
\includegraphics[scale=0.62]{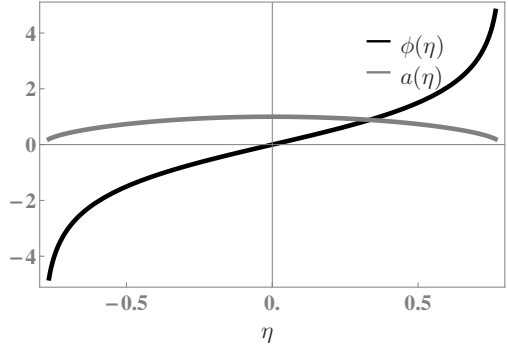}
\caption{This plot shows the behavior of $\phi(\eta)$ \eqref{phiz_sol_k1} and $a(\eta)$ \eqref{az_sol_k1}
as a function of $\eta$ for $k=1$, $a_0=c=1$ and $\phi_0=c_0=0$.}
\label{fig_K1}
\end{figure}

\paragraph{\textbf{$k=1:$}}
In this case the warp factor $a(\eta)$ found from \eqref{s1} reads
\beq
a(\eta)=a_0\sqrt{\cos(2\eta/c+c_0)},     \label{az_sol_k1}
\eeq
where $a_0$ and $c_0$ are integration constants. The above solution is applicable in the region
$|\eta+c c_0/2| < c\pi/4 $. With this expression for $a(\eta)$, we use Eq.~\eqref{s2} to find  the
scalar field $\phi(\eta)$:
\beq
\phi(\eta)=\sqrt\frac32 \ln\left(\frac{1+\tan(\eta/c+c_0/2)}{1-\tan(\eta/c+c_0/2)}\right)+\phi_0,
\label{phiz_sol_k1}
\eeq
where $\phi_0$ is an integration constant. Then the scalar potential
is given by
\beq
V(\phi)=\frac{3d^2}{2c^2}\left[3-\cosh\left(\sqrt\frac83(\phi-\phi_0)\right)\right].
\label{potential_phiz_k1}
\eeq
It is important to note that these solutions for $k=1$ case have a singularity at
$ \eta = ( \pm \pi  -2 c_0)c/4$.
Figure \ref{fig_K1} illustrates the behavior of the
scalar field $\phi(\eta)$ and the warp factor $a(\eta)$.

\paragraph{\textbf{$k=-1:$}}
In this case \eqref{s1} yields
\beq
a(\eta)=a_0\sqrt{\cosh(2\eta/c+c_0)},     \label{az_sol_k-1}
\eeq
where $a_0$ and $c_0$ are integration constants. In this case,
however, there are no real solutions for $\phi'(\eta)$, so we will not consider this possibility further.


\subsubsection{Superpotential method}
\label{Superpotential method}

It is instructive to develop an analogue of the superpotential method for the time dependent
scalar field in 5D warped space-time. We define the following quantities,
\bea
\frac{a^\p}{a}&\equiv &-\frac{1}{3}W(\phi),\lsp   \frac{\dot a}{a}\equiv -\frac{1}{3}H(\phi),
\label{super_potential_t1}
\eea
where $W(\phi)$ and $H(\phi)$ are functions of $\phi(t,y)$. In the above equation and in the following, unless otherwise stated,
we return to $t$ and $y$ derivatives by a dot and a prime
respectively.
With the above definitions we find
from the $05$ component of the Einstein equations \eqref{ein_05},
\bea
\frac{\partial W(\phi)}{\partial\phi}&=&\phi^\p,\lsp   \frac{\partial H(\phi)}{\partial\phi}=\dot \phi.
\label{super_potential_t2}
\eea
Now, if we re-express the $55$ (or $00$) component of the Einstein equations in terms of the
superpotential variables $W(\phi)$ and $H(\phi)$ through Eqs. \eqref{super_potential_t1} and
\eqref{super_potential_t2} we get the potential $V(\phi)$ as,
\begin{align}
V(\phi)=&\frac{1}{2}\left( \frac{\partial W(\phi)}{\partial \phi}\right)^{2}-\frac{2}{3}W(\phi)^{2}-\frac{1}{a^2}\left(\frac{1}{2}\left( \frac{\partial H(\phi)}{\partial \phi}\right)^{2}-\frac{1}{3}H(\phi)^{2}-3k\right).
\label{potential_t1}
\end{align}
The $ij$ component of the Einstein equation \eqref{ein_ij}
then gives
\begin{equation}
\frac{1}{2}\left( \frac{\partial H(\phi)}{\partial \phi}\right)^{2}-\frac{1}{3}H(\phi)^{2}-3k=0.
\label{H_constraint}
\end{equation}
For the $k=0$ case this gives
\begin{equation}
\frac{\partial H(\phi)}{\partial \phi}\frac{1}{H(\phi)}=\pm\sqrt{\frac{2}{3}}.
\label{H_constraint1}
\end{equation}
with solution
\begin{equation}
H(\phi)=H_0e^{\pm\sqrt{\frac23}\phi}.
\label{H_sol}
\end{equation}
where, $H_0\equiv H(0)$ is a constant of integration~\footnote{Notice that
$H(\phi)=H_0e^{\pm\sqrt{\frac23}\vert\phi\vert}$ is also a solution of \eqref{H_constraint}
even though the first derivative of $\phi$ is discontinuous which suggest that the solutions
will require presence of a singular brane. We will not consider such solution in this study.}.
It is important to note that the superpotentials $W(\phi)$ and $H(\phi)$ are related
to each other since from \eqref{super_potential_t2} one obtains
\bea
\frac{\partial^2 W(\phi)}{\partial\phi^2}\dot\phi&=&\dot\phi^\p, \lsp
\frac{\partial^2 H(\phi)}{\partial\phi^2}\phi^\p=\dot \phi^\p,    \label{super_potential_t4}
\eea
which implies, along with Eq. \eqref{super_potential_t2}, that,
\bea
\frac{\partial^2 W(\phi)}{\partial\phi^2}\frac{\partial H(\phi)}{\partial\phi}&=&
\frac{\partial^2 H(\phi)}{\partial\phi^2}\frac{\partial W(\phi)}{\partial\phi}.
\label{super_potential_W0}
\eea
Hence,
\bea
 W(\phi)&=&A_0 H(\phi)+W_0,   \label{super_potential_W}
\eea
where $A_0$ and $W_0$ are constants on integration.

In order to determine $\phi$ we use \eqref{H_constraint} together with \eqref{super_potential_t2}
to obtain,
\begin{equation}
\dot\phi=\pm\sqrt{\frac{2}{3}}H_0e^{\pm\sqrt{\frac23}\phi}.
\label{phi_dot}
\end{equation}
On the other hand, from \eqref{super_potential_t2} and \eqref{super_potential_W} we find,
\beq
\phi^\p
=\pm\sqrt{\frac23}A_0H_0e^{\pm\sqrt{\frac23}\phi}.   \label{phi_prime}
\eeq
Therefore from Eqs. \eqref{phi_dot} and \eqref{phi_prime}, we have,
\beq
\dot\phi(t,y)=\frac{1}{A_0}\phi^\p(t,y),   \label{phi_dp}
\eeq
which implies that, as claimed previously,
 $\phi$ can depend on $t$ and $y$ only
through $ \eta $ (with
$ d = A_0 \, c $). For simplicity, hereafter we choose $c=d=1$.
Then from Eq. \eqref{phi_dot} we obtain
\beq
\frac{d\phi(\eta)}{d\eta}=\pm\sqrt{\frac23}H_0e^{\pm\sqrt{\frac23} \phi(\eta)}.  \label{Phi_z}
\eeq
with solution
\beq
\phi(\eta)= \mp\sqrt{\frac32} \ln\left(-\frac23 H_0\eta + e^{\mp\sqrt\frac23\phi_0}\right), \label{Phi_solution}
\eeq
where $\phi_0$ is an integration constant. Note that the above solution is valid only for
$-\frac23H_0\eta+e^{\sqrt\frac23\phi_0}>0$ and there is a singularity at $-\frac23H_0\eta+e^{\sqrt\frac23\phi_0}=0$.
Also one can see that for the choice $H_0=-3b_0$ and $\phi_0=0$ the above result for $\phi(\eta)$
matches the one obtained in \eqref{phiz_sol}.

In order to determine the warp factor $a$ we use Eq.~\eqref{super_potential_t1}:
\begin{equation}
\frac{\dot a}{a}\equiv -\frac{1}{3}H(\phi)=\mp\sqrt{\frac{1}{6}}\frac{\partial H(\phi)}{\partial \phi}=\mp\sqrt{\frac{1}{6}}\dot \phi,
\label{dota_a}
\end{equation}
which can be solved to obtain $a(t,y)$ as,
\begin{equation}
a(t,y)=a(t_0,y)e^{\mp\sqrt{\frac{1}{6}}(\phi(t,y)-\phi(t_0,y))},
\label{a_solution}
\end{equation}
where $a(t_0,y)$ and $\phi(t_0,y)$ are functions of $y$ at
the constant time slice $t_0$.  $a(t_0,y)$
can be found by substituting the above expression for  $a(t,y)$ into the first equation in
Eq.~\eqref{super_potential_t1}; we then find
\begin{equation}
a(t_0,y)=a(t_0,y_0)e^{\mp\sqrt{\frac{1}{6}}(\phi(t_0,y)-\phi(t_0,y_0))-\frac13 W_0 y},
\label{a0_solution}
\end{equation}
inserting this in Eq. \eqref{a_solution} we find
\begin{equation}
a(t,y)=a(t_0,y_0)e^{\mp\sqrt{\frac{1}{6}}(\phi(t,y)-\phi(t_0,y_0))-\frac13 W_0 y},
\label{a_solution1}
\end{equation}
where $a(t_0,y_0)$ and $\phi(t_0,y_0)$ are constants. Since $\phi(t,y)=\phi(\eta)$; then,
for $\phi(t_0,y_0)\equiv\phi_0=0$, $W_0=0$ and $H_0=-3b_0$, we recover the result
that $a(t,y)\equiv a(\eta)$ as in Eq. \eqref{az_sol}.
\beq
a(\eta)=a_0\left(1+2b_0\eta\right)^{1/2}.   \label{az_sol_1}
\eeq

Since $W(\phi)$ has been found we can determine the potential $V(\phi)$ directly from
\eqref{potential_t1}
\beq
V(\phi)=-\frac13\left( A_0 H_0 e^{\pm \sqrt\frac23 \phi} +2W_0\right)^2 + \frac23 W_0^2.
\label{potential_t2}
\eeq
We recover the result for $V(\phi)$ as in Eq. \eqref{potential_phi0} with $H_0=-3b_0$, $A_0=1$, $W_0=0$
and the lower sign (minus sign) in the exponent.

Similarly one can reproduce all the results obtained in Sec.~\ref {Time dependent thick brane solutions}
adopting the superpotential method for non-zero $k$ values. Therefore
we can conclude that the superpotential method is equivalent to the assumption that
$\phi$ and $a$ depend on $t$ and $y$ only through $\eta=c t + d y$. The main advantage the this method
this that it reduces the second order differential equations into to first order equations which are
much easier to solve analytically.

\section{Conclusions}
\label{conclusions}

In this work we have analyzed a 5D scenario with a scalar field in the presence of gravity. We
have found solutions of the Einstein equations for the case of time-independent scalar field
assuming a conformal form of the 4D metric. Both the evolution of the scale factor, its extra-dimensional
shape and the profile of scalar field were discussed and determined for different values of spacial
curvature $k=0,\pm 1$. Also for the time-dependent scalar field $\phi(t,y)$ and 4D conformal metric,
analytic solutions were obtained in certain cases.
We have also formulated a superpotential method for $t$- and $y$-dependent profiles of the scalar field.
For the solution which has been found
both the scalar filed $\phi$ and the scale factor $a$ depend on time $t$ and $y$ only
through $\eta=ct+dy$, where $c$ and $d$ are constants.

\begin{acknowledgments}
BG thanks Jacek Pawelczyk for useful remarks. AA and BG are grateful to the NORDITA Program ``Beyond the LHC''
for hospitality where some part of this work is done. This work has been supported in part by the
National Science Centre (Poland) as a research project, decision no DEC-2011/01/B/ST2/00438.
AA acknowledges financial support from the Foundation for Polish Science International PhD Projects
Programme co-financed by the EU European Regional Development Fund.
\end{acknowledgments}

\appendix

\section{5D dilaton-like solutions}
\label{5D Dilaton-like solutions}
Here we are going to consider a stationary setup defined by the action \eqref{action} with the metric ansatz \eqref{metric}
with $a(t,y)\equiv a(y)$ and $\phi(t,y)=\phi(y)$.
The resulting Einstein equations and the equation of motion for $\phi$ (with $4M_\ast^{3}=1$)
are as follows~\footnote{In this case $k=0$ is required by the Einstein equations.}
\begin{align}
6\frac{a^{\prime2}}{a^{2}}&=\frac{1}{2}(\phi^{\prime})^{2}-V(\phi),\label{eom01_A}\\
3\frac{a^{\p\p}}{a}+3\frac{a^{\prime2}}{a^{2}}&=-\frac{1}{2}(\phi^{\prime})^{2}-V(\phi),\label{eom02_A}\\
\phi^{\prime\prime}+4\frac{a^{\p}}{a}\phi^{\prime}&-\frac{dV}{d\phi}=0, \label{eom03_A}
\end{align}
where a  prime denotes a $y$ derivative.
We assume that the scalar potential $V(\phi)$ can be expressed
in terms of the superpotential $W(\phi)$ as \cite{DeWolfe:1999cp,Ahmed:2012nh},
\begin{equation}
V(\phi)=\frac{1}{2}\left( \frac{\partial W(\phi)}{\partial \phi}\right)^{2}-\frac{2}{3}W(\phi)^{2},
\label{potential_A}
\end{equation}
where $W(\phi)$ satisfies the following relations,
\begin{equation}
 \phi^{\prime}=-\frac{\partial W(\phi)}{\partial \phi} \hspace{1cm}\text{and}\hspace{1cm} \frac{a^{\p}}{a}=\frac{1}{3} W(\phi).
 \label{super_potential_A}
\end{equation}

In addition we focus on the following form of {\it dilatonic superpotential} $W(\phi)$
\beq
W(\phi)=W_0 e^{\frac{\epsilon}{2}\phi},   \label{superpotential_A}
\eeq
where $W_0$ and $\epsilon$ are constants and as we will see the different values of $\epsilon$ will correspond
to a class of different solutions. The resulting scalar potential $V(\phi)$ is,
\beq
V(\phi)=W_0^2\left(\frac{\epsilon^2}{8}-\frac23\right)e^{\epsilon\phi}.  \label{potential_A1}
\eeq

The scalar field $\phi(y)$ and the warped function $A(y)$ obtained from Eq. \eqref{super_potential_A} are
\begin{align}
\phi(y)&=-\frac{2}{\epsilon}\ln\left(1+\frac{\epsilon^2}{4}W_0 y\right),     \label{phi_A}\\
a(y)&=a_0\left(1+\frac{\epsilon^2}{4}W_0 y\right)^\frac{4}{3\epsilon^2}.   \label{warp_A}
\end{align}
The above result \eqref{potential_A1}-\eqref{phi_A} and \eqref{warp_A} represents a class of solutions
parameterized by $\epsilon$. For $\epsilon=\pm \sqrt\frac83$, we recover our results found in
Eqs. \eqref{az_sol}-\eqref{phiz_sol} and \eqref{potential_phi0} for $W_0=3b_0$.
Whereas, for $\epsilon=\pm \sqrt\frac43$ we recover the linear dilaton solution discussed by Antoniadis et al. \cite{Antoniadis:2011qw}.
It is instructive to notice that the metric given by Eq. (11) of Ref. \cite{Antoniadis:2011qw}
coincides with \eqref{warp_A} for $\alpha=W_0$.

\providecommand{\href}[2]{#2}\begingroup\raggedright\endgroup

\end{document}